\documentclass[twocolumn,showpacs,amsmath,amssymb]{revtex4}
\usepackage{epsfig,psfrag}
\usepackage{dcolumn}
\usepackage{bm}
\usepackage{graphicx}
\usepackage{color}
\newcommand{\beq}{\begin{equation}}
\newcommand{\eeq}{\end{equation}}

\newcommand{\vrr}{{\vec{r}}}

\def\ve{{\varepsilon}}

\def\bM{{\mathbf M}}

\def\br{{\mathbf r}}

\def\bk{{\mathbf k}}

\def\a{{\alpha}}
\def\b{{\beta}}
\def\g{{\gamma}}
\def\d{{\delta}}

\newcommand{\sigmab}{\mbox{\boldmath $\sigma $}}

\newcommand{\vq}{{\vec{q}}}

\newcommand{\p}{{\partial}}

\def\half{{1\over2}}

\def\eqa{\begin{eqnarray}}
\def\eea{\end{eqnarray}}
\def\beqr{\begin{eqnarray}}
\def\eeqr{\end{eqnarray}}
\begin{document}

\title{ Ballistic Quantum Dots with Disorder and Interactions: A numerical study on the Robnik-Berry billiard}
\author{    Ganpathy Murthy$^1$, R. Shankar$^2$, and Harsh Mathur$^3$}
\affiliation{$^1$Department of Physics and Astronomy, University of Kentucky, Lexington KY 40506-0055\\
$^2$ Department of Physics, Yale University, New Haven CT 06520\\ $^3$
Physics Department, Case Western Reserve University, Cleveland, OH
44106-7079}
\date{\today}
\begin{abstract}
{ In previous work we have found a regime in ballistic quantum
dots where interelectron interactions can be treated
asymptotically exactly as the Thouless number $g$ of the dot
becomes very large. However, this work depends on some assumptions
concerning the renormalization group and various properties of the
dot obeying Random Matrix Theory predictions at scales of the
order of the Thouless energy. In this work we test the validity of
those assumptions by considering a particular ballistic dot, the
Robnik-Berry billiard, numerically. We find that almost all of our
predictions based on the earlier work are borne out, with the
exception of fluctuations of certain matrix elements of
interaction operators. We conclude that, at least in the
Robnik-Berry billiard, one can trust the results of our previous
work at a qualitative and semi-quantitative level. }
\end{abstract}
\vskip 1cm 
\maketitle

\section{Introduction}
The quest for a satisfactory theory of quantum dots is driven not only
by their obvious importance as mesoscopic devices revealed by a series
of groundbreaking experiments\cite{recent-expts}, but also by their
challenge as a unique confluence of disorder, interactions and
finite-size effects\cite{reviews}. For weak interactions, the
Universal Hamiltonian\cite{H_U,univ-ham} (UH) provides a satisfactory
description. For ballistic/chaotic quantum dots, we have
espoused\cite{mm,qd-ms1,longpaper} an approach based on the fermionic
Renormalization Group\cite{rg-shankar} (RG), $1/N$ expansions and the
fact that energy eigenstates around the Fermi energy in disordered
systems ought to be described by Random Matrix Theory
(RMT)\cite{alt1,RMT}. Our approach not only explains the UH as a fixed
point of the RG but also describes the physics outside its basin of
attraction. It predicts a phase transition at strong coupling and
allows a fairly detailed study\cite{longpaper} of the new phase and
the quantum critical region\cite{critical-fan} separating it from that
governed by the UH.

Our results, however, were predicated on a variety of RMT and RG
assumptions. To test our assumptions and the conclusions deduced from
them, we performed a detailed numerical study on a ballistic but
chaotic billiard (the Robnik-Berry billiard\cite{robnik-berry}) and we
report our findings here. Our expectations are borne out, with one
notable exception.

We recall our strategy and assumptions briefly so that the reader may
see in advance what sort of ideas are put to test in our study. In the
primordial problem of interest to us one has electrons confined to a
ballistic dot of size $L$, with no impurities inside, and edges so
irregular that classical motion is chaotic. The electrons experience
the Coulomb interaction. In momentum space, all momenta within the
bandwidth (of order $k_F$, the Fermi momentum) exist.  The
semiclassical ergodicization time for an electron within the dot is a
few bounces, or $\tau_{erg}\simeq L/v_F$. By the uncertainty principle
this leads to an important energy scale, the Thouless energy
$E_T\simeq\hbar v_F/L$ which has a dual significance. First, it
controls the dimensionless conductance of a dot strongly coupled to
leads, as follows. Since the transport through the dot takes place in
a time such that energy is uncertain by an amount $E_T$, all single
particle states that fit into this band will each contribute a unit of
dimensionless conductance. If the average single-particle level
spacing is $\delta$, then the dimensionless conductance is
$g=E_T/\delta$. Second, in the other limit of dots very weakly coupled
to leads (which we focus on in this work), the Thouless band of width
$E_T$ centered on $E_F$, marks the scale deep within which RMT should
apply to the energies and eigenfunctions\cite{alt1}. In this context
$g$ is better denoted the Thouless number.

Since we are only interested in a narrow band of energies of width
$E_T <<E_F$, the first step in the program\cite{mm} is to use the RG
for fermions\cite{rg-shankar} to get an effective low energy theory by
eliminating all momentum states outside $E_T$.  Should we worry that
we are not eliminating exact single particle eigenstates (labelled
here by $\a$)? No, because the disorder due to the boundaries will mix
momentum states at roughly the same energy, and it does not matter
whether we eliminate momentum states within any annulus of energy
thickness $E_T$ or the single particle states they evolve
into. Indeed, even the mixing within $E_T$ is due to the fact that
momentum itself not well defined in a finite dot, a point we will
elaborate on shortly. However, once we come down to within $E_T$
of $E_F$, we cannot eliminate the remaining states in one shot since
it is the flow of couplings {\em within} this band that is all
important in the RG.

\begin{figure}[ht]
\includegraphics*[width=2.4in,angle=0]{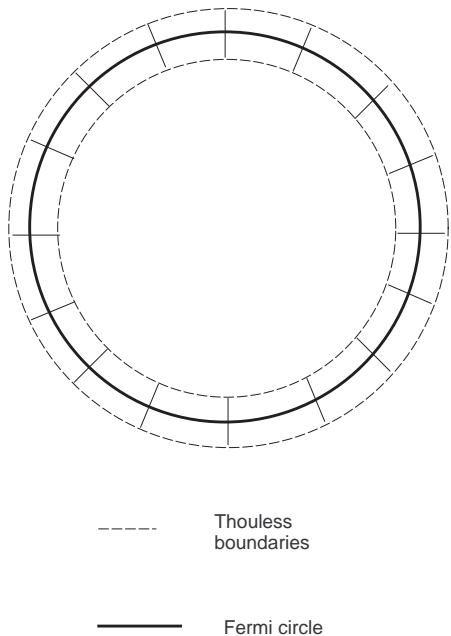}
\caption{The $g$ wheel-of-fortune states in the Thouless band.
They are packets in $\bk$ space with average momenta equally
spaced on the Fermi circle.} \label{wof}
\end{figure}

 Now it is
known\cite{rg-shankar} that the clean system RG (justified above)
leads to Landau's Fermi liquid interaction\cite{agd}
\beq
V=\sum_{\bk \bk'} F(\theta_{\bk}-\theta_{\bk'})\delta n(\bk
)\delta n(\bk' )
\eeq
at an energy scale $E_L$ which is small compared
to $E_F$. But since $E_L$ is a bulk scale it can always be made larger
than $E_T$ which vanishes as $L\to
\infty$. Thus Murthy and Mathur\cite{mm} perform their RG on the
hamiltonian (focussing on the spinless case for simplicity)
\beq H=\sum_{\a}c^{\dag}_{\a}c_{\a}\varepsilon_{\a}+
\sum\limits_{\a \b \g \d}V_{\a \b \g
\d}c^{\dag}_{\a}c^{\dag}_{\b}c_{\g}c_{\d}
\eeq
where
\beqr
 V_{\a \b
\g \d} =&{1\over4} \sum\limits_{ \bk \bk'}F(\theta_{\bk}-\theta_{\bk'})(\phi^{*}_{\a}(\bk )\phi^*_{\b}(\bk')-\phi^*_{\b}(\bk)\phi^*_{\a}(\bk'))\nonumber\\
&\times(\phi_\g(\bk')\phi_\d(\bk)-\phi_\d(\bk')\phi_\g(\bk)) \label{wof1}
\eeqr
is simply the Landau interaction written in the basis of exact
eigenstates, a statement that needs some elaboration. In usual RMT
treatments,   $\phi_{\a}(\bk )$ is the exact eigenstate $\a$ written
in the infinite dimensional basis of all momentum states. In our
version which uses  the RG to reduce the Hilbert space, the states
labeled by $\bk$ are approximate momentum states with an uncertainty
$\delta k\simeq 1/L$ in both directions. The number of such wave
packets that fit into an annulus of radius $k_F$ and thickness
$E_T/v_F$ is ${\mathcal{O} } (k_F L)=g$. We call them the
Wheel-of-Fortune (WOF) states, see Figure (\ref{wof}). One way to
construct such packets is to pick $g$ plane waves of equally spaced
momenta on the Fermi circle and to chop them off at the edges of the
dot to respect the boundary conditions. This is what we mean by
$\bk$ in $\phi_{\a}(\bk )=\langle \bk |\a\rangle$.

 We are now ready to state our two assumptions.\\
{\em {\bf Assumption 1}: We assume that the $g$ approximate
momentum states generated as above form a complete basis for the
$g$ exact eigenstates in the Thouless shell}.

{\bf Assumption II}: The energy eigenvalues $\varepsilon_{\a}$
obey RMT statistics as do the wavefunctions. For example we assume
that the ensemble averages (denoted by $\langle \rangle$) obey
\beq \langle \phi^{*}_{\mu}(\bk_1 )\phi^{}_{\mu}(\bk_2
)\phi^{*}_{\nu}(\bk_3 )\phi^{}_{\nu}(\bk_4) \rangle
={\delta_{12}\delta_{34}\over g^2} +O(1/g^3) \label{4pt}\eeq

To some, Assumption I seems remarkable -- How can we furnish {\em in
advance, independent of the dot shape} a basis of $g$ states for
expanding the $g$ exact eigenstates within $E_T$? After all, these
eigenstates are supposed to resemble those of a random matrix.  The
point is that no matter how chaotic the dot, it can only mix states at
the same energy. While this sounds like Berry's
ansatz\cite{berry-ansatz} it is somewhat different in both scope and
content: Berry's ansatz states that every exact eigenstate $\a$ can be
expanded in terms of an {\it infinite number} of $\bk$ states (with
the same energy $\epsilon_{\a}=\bk^2/2m$)in the bulk, while we claim
that only $g$ of WOF states are needed. Secondly, we claim that the
{\it same} $g$ WOF states can be used to describe all the states
within the Thouless band.

In this work we will show that these states are nearly orthonormal and
that the exact state right in the middle of the band has 99.9\%
overlap with the WOF states. The success of this extension of Berry's
conjecture to a finite dot exceeds our expectations in this
regard. However, we find that the $g$ WOF states become less effective
at describing states as we move away from $E_F$: the overlap drops to
50\% at $E_{\a}=E_F\pm E_T/2$. This
prepares us for the possibility that nonuniversal quantities may be
quantitatively inaccurate in our approach.

As for Assumption II, we have verified RMT behavior for the
eigenvalues (as have others before
us\cite{stone-bruus,alhassid-lewenkopf}) but not the eigenfunctions.
What we did instead was to see what extent the solution of a specific
dot resembled the picture we drew based on these two assumptions. We
begin by describing how one starts from Eqn. (\ref{wof1}), which
describes the effective hamiltonian, and use our two assumptions with
large-$N$ ideas to make our predictions\cite{qd-ms1,longpaper}. These
predictions are asymptotically exact as $g\to\infty$.

First one expands the Landau function as 
\beq F(\theta ) = \sum_m u_m
e^{im\theta}.
\eeq 
Barring accidents, the phase transition occurs in one channel with
some $m$ (recall superconductivity). This allows us to focus on a
single $u_m\ne 0$, ignoring all others. Then we carry out a
Hubbard-Stratovich transformation on the interaction using a
collective field $\sigmab$. We then formally integrate out the
fermions and get an effective action $S(\sigmab )$ for $\sigmab$.  In
this process we make use of assumption II. The action in terms of
$\sigmab$ is obtained by summing one loop Feynman diagrams with
varying numbers of external $\sigmab$'s connected to a single fermion
line running around the loop. Each diagram is a sum over fermion
energy denominators multiplying products of a string of $\phi_{\a}(\bk
)$'s.  We are able to show that these products may be replaced by
their ensemble averages in the large $g$ limit. In other words the sum
over so many terms in each diagrams leads to self-averaging. For the
averages we use relations like Eqn.  (\ref{4pt}). When this is done,
the effective action can be cast into a form which has a $g^2$ in
front of it\cite{qd-ms1,longpaper}, so that the saddle point gives
exact answers as $g \to \infty$. 

At this point let us collect all the results and predictions of the
RMT + large-$N$ theory\cite{longpaper} with a view to comparing them
with similar results without using Assumptions I and II on the
Robnik-Berry billiard.
\begin{itemize}
\item{}  In the large-$g$ limit there is a sharp transition to a
phase in which $\sigmab$ acquires a vacuum expectation value. The
critical value of $u_m$ in our approximation turns out to be
$-1/\log{2}$ in the spinless case and $-1/2\log{2}$ in the spinful
case. The true critical value is most likely to be the bulk value
$-2$ (spinless) or $-1$ (spinful), as has been found in an
explicitly solvable model by Adam, Brouwer, and Sharma
recently\cite{nowindow}. This is an example of the nonuniversal
quantity alluded to earlier, that we cannot predict exactly in our
approach even as $g\to \infty$.

\item{}  For finite $g$, instead of a sharp phase transition, there is
a crossover from the weak-coupling regime through a quantum critical
regime to a strong-coupling regime. Due to the explicit
symmetry-breaking at order $1/g$, there is always some nonzero order
parameter, which increases to a number of order $g$ (in the
normalization we use here, which is different from that of
ref.\cite{longpaper}) in the strong-coupling regime.

\item{} For symmetry-breaking in  odd angular momentum channels there are
two exactly degenerate minima for every sample arising from
time-reversal invariance.

\item{} The ground state energy at the minimum in the strong-coupling
regime is lower than that in the weak-coupling regime by a number of
order $g^2\d$.

\item{} The effective potential landscape in the strong-coupling
regime is in the approximate shape of a Mexican Hat, with the ripples
at the bottom of the hat being of order $g\d$.

\item{} In the quantum-critical and strong-coupling regimes,
even low-energy quasiparticles acquire large widths given on average by
\beq
\Gamma(\varepsilon)\approx {\d\over\pi}\log(\varepsilon/\d)
\eeq
\end{itemize}

We found that most of these predictions are verified by our
numerical results on the Robnik-Berry billiard, except that the
ripples at the bottom of the Mexican Hat turn out to be much
larger than expected for the $m=2$ Landau interaction channel.

\section{The Robnik-Berry billiard}
In this section we will describe how the dot is chosen and how the
single particle energy levels $\varepsilon_{\a}$ and eigenfunctions
$\phi_{\a}(\br )$ are determined. We use a trick invented by Robnik
and Berry\cite{robnik-berry} and elaborated upon by
Stone and Bruus\cite{stone-bruus}. Consider a unit circle $|z|=1$ in the complex
plane of $z=x+iy$. The analytic function
\beq
w(z) = {z+bz^2+cz^3e^{i\chi}\over \sqrt{1+2b^2+3c^2}}
\label{conformal-map}\eeq
defines a map
under which the unit circle in $z$ gets mapped into a new shape in
$w$, which will be our dot. The shape of the dot can be varied by
varying the parameters $b,\ c, \ \mbox{and}\ \chi$. The denominator
ensures that the billiard has the same area ($\pi$) as the unit disc.
The wavefunction $\phi_{\a}(w,\bar{w}) =\phi_{\a}(u, v) $ is required
to vanish at the boundary and obey
\begin{eqnarray}
&-&\left( {\p^2\over \p
u^2}+{\p^2\over \p v^2}\right) \phi_{\a}(u,v)\nonumber \\
&=&-4{\p \over \p w} {\p \over \p \bar{w}}\phi_{\a}(w,\bar{w})
=\varepsilon_{\a} \phi_{\a}(w,\bar{w}).
\end{eqnarray}
(We have
chosen $\hbar = 2m =1$). If we now go the $z$ plane where the
wavefunction is $\phi_{\a}(w(z),\bar{w}(\bar{z}))$, the
Schr\"{o}dinger equation and boundary condition are
\beq
-4{\p
\over \p z} {\p \over \p
\bar{z}}\phi_{\a}(z,\bar{z})=\varepsilon_{\a}|w'(z)|^2
\phi_{\a}(z,\bar{z}) \label{se} \ \ \ \ \phi_{\a}(|z=1|)=0
\eeq
where $w'(z)=dw/dz$. This differential equation in the continuum
is next converted to a discrete matrix equation by writing
\beq
\phi_{\a}(z,\bar{z})\equiv \phi_{\a}(r,\theta)=\sum_j {1\over
\g_j}C^{\a}_{j}\psi_j(r,\theta )\label{exp}
\eeq
where
$\psi_j(r,\theta ) $ is the solution to the  free Schr\"odinger
equation in the unit disk vanishing on the boundary:
\beq
-\nabla^2 \psi_j(r,\theta )=\g^{2}_{j}\psi_j(r,\theta ).
\eeq
(That is, these are Bessel functions in $r$ times angular momentum
eigenfunctions in $\theta$. ) Feeding this expansion into Eqn.
(\ref{se}) one obtains the matrix equation \beq \sum_j
M_{ij}C^{\a}_{j}={1 \over \varepsilon_{\a}}C^{\a}_{i}\eeq where
\beq M_{ij}={1\over \g_i}\langle i||w'|^2|j\rangle {1\over \g_j}.
\eeq (Without the $1/\g_j$ in the expansion Eqn. (\ref{exp}), $M$
would not have been Hermitian. )

In practice one truncates $M$ to a finite size (we used 585
states) and expects the lower energy levels and wavefunctions to
be unaffected.

\begin{figure}[ht]
\includegraphics*[width=2.4in,angle=0]{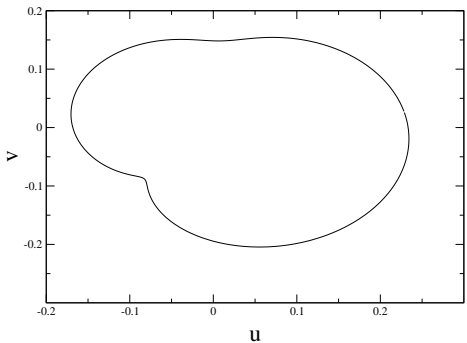}
\caption{The shape of the Robnik-Berry billiard for $b=c=0.2$, and $\chi=0.85$.}
\label{fig1}
\end{figure}

The parameters $b,c,\chi$ are chosen to lie in the range where
classical behavior is chaotic, and where quantum chaos as reflected in
the eigenvalue distribution has been established\cite{stone-bruus}. A
value we used repeatedly was $b=c=.2,\chi =.85$. A nonzero $\chi$ ensures
that the billiard has no reflection symmetry. This shape is often
called the Africa Billiard based on its resemblance to that continent,
as seen in Fig \ref{fig1} for our chosen values of parameters.

We shall refer to these eigenfunctions and eigenvalues as exact even
though they come from solving a truncated problem because we can
easily increase the accuracy by increasing the size of the truncated
Hilbert space.

\subsection{ Testing Assumptions I and II}

Our ability to solve the Schr\"odinger equation (to high accuracy)
implies in principle that we can test our two assumptions.

In the next subsection we will test Assumption I, i.e.,  see in
detail how well the WOF states serve a basis within the Thouless
band.

As for Assumption II, we and our predecessors
\cite{stone-bruus,alhassid-lewenkopf} have shown that the
eigenvalues and single eigenfunctions obey the distribution
expected by RMT for a GOE. (The ensemble is generated by varying
the parameters in $w(z)$.)

Similar information about  wavefunction correlations is not known
in the ballistic problem (despite some recent progress using
supersymmetry methods\cite{supersymm}). We did not try to do this
here since  our computing capabilities did not allow us to
generate an ensemble.

Instead we computed the fate of the interacting system without
recourse to Assumptions I and II and compare  to our predictions
based on these assumptions.

\subsection{Completeness of the WOF basis}
Let $E_F$ be the Fermi energy. Then $g\simeq \sqrt{4\pi
N}\simeq\sqrt{\pi E_F}$, which we arrive at as follows. The Fermi
circle has a circumference $2\pi K_F$ and into this will fit
$g=2\pi K_F/(2\pi /L)$ WOF states each of width $2\pi /L$ in the
tangential direction. Finally $E_F=K_{F}^{2}/2m = K_{F}^{2}$,
$N=k_F^2L^2/4\pi$ and $L=\sqrt{\mbox{Area}}=\sqrt{\pi}$.

As a test case when we picked the Fermi energy to be the 100-th
level, we found $g=37$. How well is this state $|F>$ at $E_F$
spanned by the $g$ WOF states at the Fermi energy?

\begin{figure}[ht]
\includegraphics*[width=2.4in,angle=0]{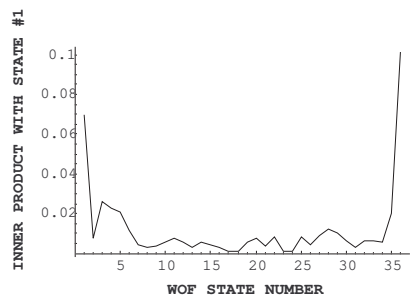}
\caption{The absolute value of $\langle n|1\rangle$, the inner
product of WOF state number 1 with the other$g-1=36$ states. }
\label{fig2a}
\end{figure}

First we first take $g$ equally spaced points $\bk_n$ on the Fermi
circle and form the WOF states \beq \psi_{WOF-n}(\br ) = {1\over
\sqrt{\pi}}e^{i\bk \cdot \br} \Theta (\mbox{dot})\eeq where $\Theta
(\mbox{dot})$ is unity inside the dot and zero outside. These states
are very close to being orthonormal. For example the overlaps of $n=1$
state (with $\bk $ along the $y$-axis) with the others as we go around
the circle is shown in Fig. (\ref{fig2a}).

Next we ask how much of the  state $|F>$ at the Fermi energy is
contained in the WOF states. We find $\sum_{n=1}^{g}|\langle
n-WOF|F\rangle |^2 = .9993$. This is a rather remarkable result. It
says that $|F>$, which is a vector with 580 components (which was
the size of our truncated problem) can be expanded almost completely
in terms of $g=37 $ WOF states which are given in advance. In other
words as one changes the shape of the dot and works at fixed Fermi
energy, the state $|F>$ changes in a random way, but that randomness
is only in which particular combination of WOF states describes it,
not in the completeness of the WOF basis.

\begin{figure}[ht]
\includegraphics*[width=2.4in,angle=0]{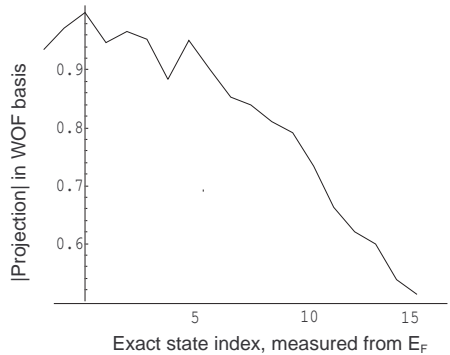}
\caption{The norm of the projection of the $n^{th}$ exact eigenstate from $E_F$ 
on to the subspace spanned by the WOF states. It can be seen that more
and more of the exact eigenstate lies outside this subspace as one
moves away from $E_F$. } \label{fig2b}
\end{figure}

While this is very satisfactory we need more to implement our scheme:
we need to be expand all $g$ states in the WOF basis. Here we find
that as we move off the center of the Thouless band, the fractional
norm captured by the WOF basis drops. In a typical case, with $g=37$,
there are roughly 12 states (one third of $g$) where the number lies
above 95\%. At band edge this drops to 50\%, as shown in Fig
(\ref{fig2b}).  Thus there is inevitably some error in transcribing
the Landau interaction written in terms of the WOF states labelled by
$\bk$ into the basis of $g$ exact eigenstates labelled by $\a$. This
just means that the location of the critical point will not be
correctly predicted by our RMT based analysis, as pointed out recently
by Adam, Brouwer, and Sharma\cite{nowindow}.

This concludes our (partial) test of Assumptions I and II. We turn
to a comparison of our results based on these assumptions with a
direct solution of the problem with no recourse to the
assumptions.

\subsection{Hartree-Fock  solution of  interacting problem}

How can  the knowledge of the "exact" eigenfunctions and
eigenvalues in the billiard  help in the solution of the problem
with interactions? The tactic will be illustrated in schematic
form first. Suppose we have a four-Fermi  interaction added to a
free hamiltonian which in first-quantization is given by some
differential operator $H_0$. Then the path integral becomes
\beq Z=\int d\psi d\bar{\psi} e^{S}
\eeq
where
\beq S= \int
d\tau \left[ \bar{\psi} \left( i\partial_{\tau} - H_0 \right) \psi
+{u\over 2}  (\bar{\psi}\psi )^2\right] .
\eeq
Using a Hubbard-Stratanovich transformation we can write
\beq Z=\int d\psi d\bar{\psi} d\sigma
e^{S}
\eeq
where
\beq S= \int d\tau \left[ \bar{\psi} \left[
i\partial_{\tau} - H_0-\sigma  \right] \psi -{\sigma^2\over 2u}
\right] .
\eeq
If the fermions are integrated out we will get an effective action
$S_{eff}(\sigma )$. {\em To find the minimum we need just the
action for static $\sigma$. } In this case it is clear that
\beq
\int d\psi
d\bar{\psi}\exp \left[ \int d\tau  \bar{\psi} \left[
i\partial_{\tau} - H_0+\sigma \right] \psi \right]=
e^{-E_0(\sigma) T}
\eeq
where $T\to \infty$ is the length of the imaginary time $\tau$- axis
and $E_0(\sigma ) $ is the ground state energy of $\psi^{\dag}
(H_0+\sigma )\psi$. To find $E_0(\sigma )$ one simply solves for the
single particle levels of $ (H_0+\sigma )$ and fills up the ones with
negative energy. The effective action for static configurations, which
is also the effective potential, is
\beq
V_{eff}=E_0(\sigma )+{\sigma^2 \over 2u}.
\label{effpot}\eeq

At this point we have a mean-field theory. We still need to
justify its use by showing that fluctuations of the collective
field $\sigmab$ around its minimum are small. In our previous
work, based on Assumptions I and II we showed that the
fluctuations were indeed small in the limit of large $g$, since
the $g^2$ in front of the actions  limits fluctuations. In the
billiard we will justify the mean field similarly, based on the
depth and curvature of the minimum.

When the  Landau  interaction is factorized,  the hamiltonian
whose ground state  gives us $E_0(\sigma )$ is
\beq
\sum_{\a \b} \psi^{\dag}_{\a} (\delta_{\a \b}{\large
\varepsilon}_{\b}+\sigmab\cdot \bM_{\a \b})\psi_{\b}
\label{ham-to-diag}\eeq
where, for the case $m=1$, for example,
\beq \bM_{\a \b}=\sum_{\bk} \phi^{*}_{\a}(\bk )\phi_{\b}(\bk )
{\bk \over k}
\eeq
and $\a , \b, \bk$ are not restricted to the Thouless band. This is
because we want to solve the problem without any of the assumptions
that led to the effective low energy theory within the Thouless
band. Note that $\sigmab$ has two components, because the Landau
interaction associated with $u_m$ has two parts:
\beq
V_{L}= {u_m\over2}\sum_{\bk \bk'} \delta n_{\bk}
\delta n_{\bk'}(\cos m \theta_{\bk} \cos m \theta_{\bk'}+ \sin m
\theta_{\bk} \sin m \theta_{\bk'}).
\eeq
Once $S_{eff}$ is known (on a grid of points in the $\sigmab$
plane) one can ask if and when the minimum moves off the origin.

So far our considerations have been fairly generic, and the Landau
interaction has been written in momentum space. However, in testing
our approach in the billiard, we will find it more convenient to
represent the Landau interaction in real space, since the
eigenfunctions are known as linear combinations of Bessel functions
whose integrals are best carried out in real space. We have carried
out calculations for two Laudau parameters corresponding to $m=1$ and
$m=2$. The $m=1$ Landau interaction is chosen to be (in
second-quantized notation)
\beqr
\half &\int d^2r \Psi^{\dag}(\vrr){1\over(2mH_0)^{1/4}}(-i{\vec\nabla}){1\over(2mH_0)^{1/4}}\Psi(\vrr) \cdot\nonumber\\
&\times\int d^2r' \Psi^{\dag}(\vrr'){1\over(2mH_0)^{1/4}}(-i{\vec\nabla}'){1\over(2mH_0)^{1/4}}\Psi(\vrr')
\eeqr
The factors of ${1\over(2mH_0)^{1/4}}$ on each side of the $\nabla$
have the effect of $1/|\bk|$ in momentum space. Since momentum does
not commute with the free Hamiltonian $H_0$, the factors have to be
placed symmetrically. Note that this corresponds only to the $\vq=0$
part of the Landau interaction. In reality, all values of $\vq$ up to
the scale $E_L/v_F$ exist in the Hamiltonian. Depending on the shape
of the dot a particular combination of them may break symmetry to give
the best energy. Still, we expect that since at large $g$ we are close
to the zero-dimensional limit, the best combination will consist
largely of very small $\vq$ parts of the Landau interaction. In any
case, the energy of the true symmetry-broken state can only be
lower than what we calculate, so what we have here is a conservative
estimate of symmetry-breaking. Similarly the $m=2$ interaction (also
at $\vq=0$) is
\beqr
\half&\int d^2r \Psi^{\dag}(\vrr){1\over(2mH_0)^{1/2}}(\nabla_x^2-\nabla_y^2){1\over(2mH_0)^{1/2}}\Psi(\vrr) \cdot\nonumber\\
&\times\int d^2r'
\Psi^{\dag}(\vrr'){1\over(2mH_0)^{1/2}}((\nabla')_x^2-(\nabla')_y^2){1\over(2mH_0)^{1/2}}\Psi(\vrr')\nonumber\\
&+\half\int d^2r
\Psi^{\dag}(\vrr){1\over(2mH_0)^{1/2}}2\nabla_x\nabla_y{1\over(2mH_0)^{1/2}}\Psi(\vrr)
\nonumber\\
&\times\int d^2r'
\Psi^{\dag}(\vrr'){1\over(2mH_0)^{1/2}}2(\nabla')_x(\nabla')_y{1\over(2mH_0)^{1/2}}\Psi(\vrr')
\eeqr
The integrals are over $(w,\bar{w})$, but can be converted to
integrals over the disk by using the conformal mapping of Eq.
(\ref{conformal-map}). Of course, the derivative operators must
also be transformed in the process. In order to find the matrix
elements of $\bM_{\a \b}$ we had to take the matrix elements of
the above operators in the basis of exact billiard states. We
carried out the angular part of the integrals analytically, but
had to turn to numerical integration to evaluate the radial
integrals. This is a computationally  intensive calculation, but
once the matrix $\bM$ has been constructed, one simply
diagonalizes the Hamiltonian of Eq. (\ref{ham-to-diag}) for a mesh
of $\sigmab$ in the plane, adds up the energies of the lowest $N$
particles to obtain the fermionic ground state energy, and obtains
the effective potential landscape from Eq. (\ref{effpot}) for
various values of the coupling strength $u$. After this, it is a
simple matter to identify the global minimum, which gives us the
lowering of ground state energy and the value of the order
parameter as a function of $u$.

\begin{figure}[ht]
\includegraphics*[width=2.4in,angle=0]{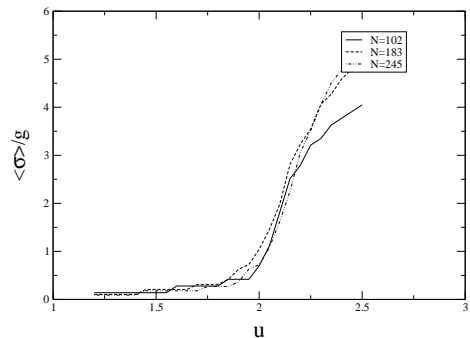}
\caption{The absolute value of the order parameter normalized by $g$
 as a function of coupling strength $u$ for three values of the number
 of particles $N$.  The fact that the curves track each other closely
 indicates that the order parameter does indeed scale like
 $g$. Furthermore, nothing discontinuous happens at the bulk critical
 coupling strength $u^*_{bulk}=2$.}
\label{fig3}
\end{figure}

Let us proceed to the results,  displayed in pictorial form. In
Fig. \ref{fig3} we show the absolute value of the order parameter,
normalized by the nominal value of $g=\sqrt{4\pi N}$, for three
values of the number of particles $N$. The bulk transition happens
at $u^*_{bulk}=2$. As can be seen, there is a nonzero order
parameter for any nonzero $u$, and it grows smoothly and
continuously as $u$ increases. Nothing discontinuous happens at
$u=2$ or even beyond, indicating that the instability does not
suddenly become first-order at the bulk value of $u^*$. Of course,
in these finite systems, the Thouless and bulk scales are related
by a factor $g/4\pi$, which is not that large (4.4 for the largest
system we considered, with $N=245$). So somewhere between $u=2.25$
and $u=2.5$ the instability seems to reach the bulk scale.
However, note that the size of systems we have considered
correspond quite closely to actual ballistic
samples\cite{recent-expts}, which typically have a few hundred
electrons. Further, the three curves seem to track each other
fairly closely, indicating that the expectation value of
$|\sigmab|$ indeed scales with $g$, as predicted by our earlier
work based on RMT assumptions.

\begin{figure}[ht]
\includegraphics*[width=2.4in,angle=0]{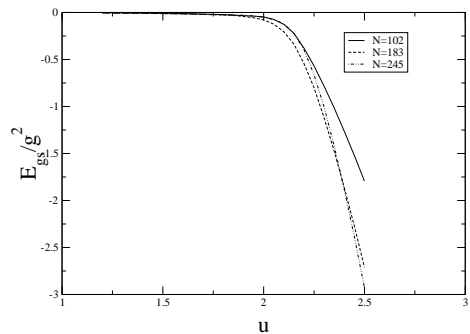}
\caption{Reduction in ground state energy normalized by $g^2$ for three values of the number of particles $N$.}
\label{fig4}
\end{figure}

In Fig. \ref{fig4} we show the corresponding reduction in ground state
energy normalized by $g^2$. Once again, the curves track each other
fairly closely, indicating that the energy reduction due to
interactions is indeed of order $g^2$, as predicted by our earlier
work.

\begin{figure}[ht]
\includegraphics*[width=2.4in,angle=0]{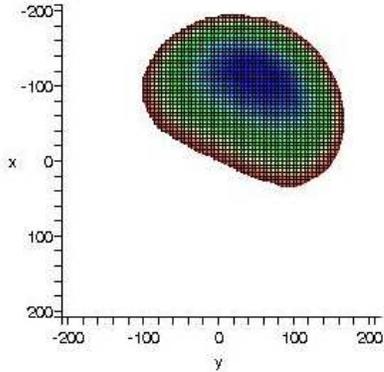}
\caption{Effective potential landscape for symmetry-breaking in the
$m=2$ channel. Instead of a Mexican Hat minimum structure with small
ripples we see an isolated minimum. The minimum does seem shallower in
the transverse direction. }
\label{fig5}
\end{figure}

In Fig. \ref{fig5} we show the effective potential landscape for
$m=2$, with $N=245$, at a value of $u=2.15$, at which the minimum is
well-developed, but the order parameter is still within the nominal
Thouless scale and has not reached the bulk scale. The RMT analysis
predicted a Mexican Hat landscape with ``small'' ripples (down by
$1/g$) in the circle of minima of the Mexican Hat. The landscape we
see bears no resemblance to this. Instead, it appears to be an
isolated minimum at a nonzero $\sigmab$. Upon close inspection it can
be seen that the minimum is shallower in the transverse direction than
in the radial direction, but this is the only indication we could find
of a (perhaps) incipient Mexican Hat structure.

\begin{figure}[ht]
\includegraphics*[width=2.4in,angle=0]{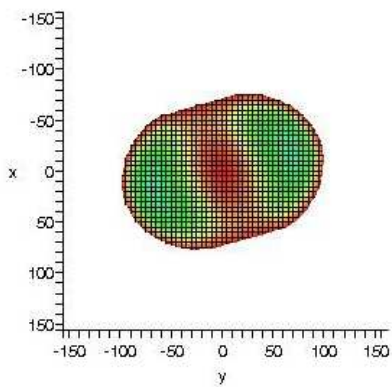}
\caption{Effective potential landscape for symmetry-breaking in the
$m=1$ channel for $N=200$ and $u=2.15$. The two exactly degenerate
minima required by time-reversal invariance can be seen, as can a
Mexican-Hat-like structure.}
\label{fig6}
\end{figure}

Fig. \ref{fig6} shows a similar effective potential landscape for
symmetry breaking in the $m=1$, channel, where the two exactly
degenerate minima expected from time-reversal invariance
considerations can be seen. The landscape also appears more
Mexican-Hat-like than in the $m=2$ case.

\begin{figure}[ht]
\includegraphics*[width=2.4in,angle=0]{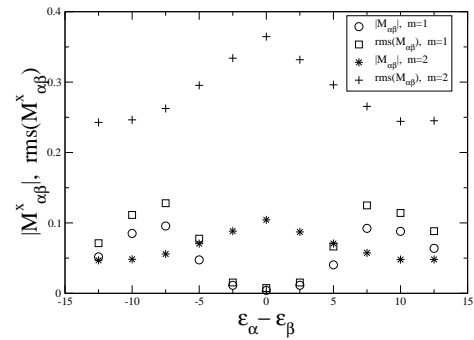}
\caption{A plot of the absolute value and the rms deviation  of the matrix elements
$M^x_{\a\b}$ from their mean absolute value as a function of
$\ve_{\a}-\ve_{\b}$ for the two cases $m=1,2$.}
\label{fig7}
\end{figure}

To trace the origin of this difference in behavior, we investigated
the average absolute value $\langle |M^i_{\a\b}|\rangle$ and the rms
deviation of the matrix elements from the mean absolute value,
$\sqrt{\langle |M^i_{\a\b}|^2\rangle-\langle |M^i_{\a\b}|\rangle^2}$
for the two cases $m=1,2$. The results for the $i=1$ (corresponding to
$\nabla_x$ for $m=1$ and $\nabla_x^2-\nabla_y^2$ for $m=2$) shown in
Fig. \ref{fig7} are an energy average for a particular billiard, with
the parameters $b=c=0.20, \d=0.85$. (We have confirmed similar
behavior of the matrix elements for other parameter values as well.)
Fig.
\ref{fig7}  shows these quantities  as a function of the energy
difference between the two states $\a$ and $\b$. There are two
features that are particularly noteworthy.

\begin{itemize}
\item There is a ``hole'' in the $m=1$ matrix element near zero
energy difference.

\item The rms deviation of the $m=2$ matrix elements from their
mean absolute value is huge. As a rough estimate, if the matrix
elements were Gaussian distributed complex numbers, the rms
deviation should be roughly half the mean modulus.
\end{itemize}

\begin{figure}[ht]
\includegraphics*[width=2.4in,angle=0]{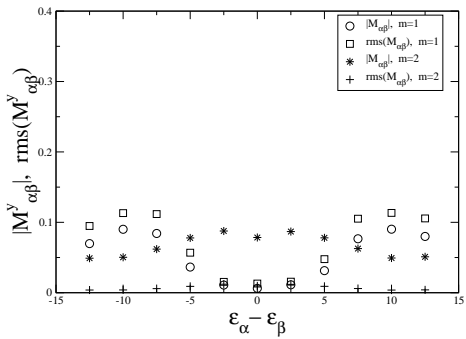}
\caption{A plot of the absolute value and the rms deviation  of the matrix elements
$M^y_{\a\b}$ from their mean absolute value as a function of
$\ve_{\a}-\ve_{\b}$ for the two cases $m=1,2$.}
\label{fig8}
\end{figure}

However, the $i=2$ component (corresponding to $\nabla_y$ for $m=1$
and $2\nabla_x\nabla_y$ for $m=2$) shows very different behavior in
Fig. \ref{fig8}. While the $m=1$ case looks similar to the $i=1$
component, the fluctuations of the $m=2$ $i=2$ component are strongly
suppressed by almost an order of magnitude below the mean.

Consider first the "hole" at $E_F$ for $m=1$.  By symmetry
considerations alone one can understand that the diagonal matrix
element $M_{\a\a}$ for $m=1$ has to be zero sample by sample in the
absence of an external magnetic field. Focusing on the $x$ component
of the order parameter
\beq
M^{x}_{\a\a}=\sum\limits_{\bk} \cos(\theta_{\bk}) \phi^*_{\a}(\bk)\phi_{\a}(\bk)
\eeq
By time-reversal invariance $\phi^*_{\a}(\bk)=\phi_{\a}(-\bk)$. Noting
that $\theta_{-\bk}=\theta_{\bk}+\pi$, and that the $\cos$ term
changes sign, one concludes that $M_{\a\a}=-M_{\a\a}=0$. The reason
the "hole'' persists for finite energy differences for the operator
${\vec p}=-i{\vec\nabla}$ can be explained by noting
that\cite{barnett-pvt} for a billiard
\beqr
{\vec p}=&im[\vrr,H]\nonumber\\
\Rightarrow (-i{\vec\nabla})_{\a\b}=&-im(\vrr)_{\a\b} (\ve_\a-\ve_\b)
\eeqr
which means that the matrix element must vanish at least linearly
with the energy difference. In fact, such ``banded'' matrix
elements have been found for many operators in ballistic
dots\cite{barnett}.

Consider next the fact that  the distribution of the matrix
elements of $M^x_{\a\b}$ for the $m=2$ case is much broader than
for the $m=1$ case, while the $M^y_{\a\b}$ matrix elements have a
very narrow distribution. The RMT answer would have the rms
deviation of $M_{\a\b}$ from the mean to be of the same order as
the mean absolute value. This seems to be roughly true for both
components of $m=1$ but grossly untrue for the $i=1$ component of
$m=2$. Since it is these mesoscopic fluctuations in $M_{\a\a}$
which determine the size of the ripples at the bottom of the
Mexican Hat in the RMT scenario, this broad distribution of
$M_{\a\b}$ seems to be the cause of the failure of the RMT
prediction that the ripples should be subdominant by $1/g$. While
it is tempting to try to explain this in relation to the shape of
the billiard (Fig. \ref{fig1}), which certainly appears to favor
an $x^2-y^2$ type of symmetry, a satisfactory explanation  of the
broad distribution of the $m=2$, $i=1$ matrix elements eludes us.

\begin{figure}[ht]
\includegraphics*[width=2.4in,angle=0]{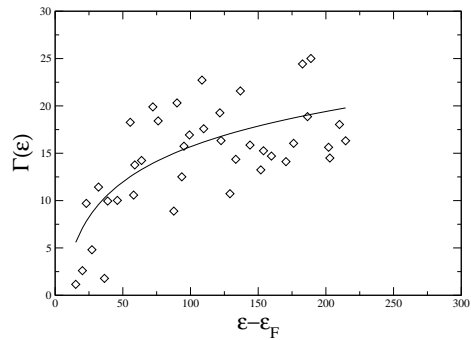}
\caption{A plot of the decay width of quasiparticles induced by their coupling to
fluctuations of the collective field $\sigmab$, for $N=245$, $m=2$,
and $u=2.15$. The solid line is the theoretical prediction from our
previous RMT-based analysis\cite{longpaper}. While the prediction does
well on average there are huge variations in the widths due to large
variations in how strongly each level couples to the collective mode.}
\label{fig9}
\end{figure}

Our knowledge of the eigenfunctions at the global minimum allow us
to compute the  effective action for time-dependent $\sigmab$ at
that minimum. Since the quasiparticles couple to this collective
field, the interaction induces a decay width for the
quasiparticles (details can be found in ref.\cite{longpaper}). In
Fig. \ref{fig9} we compare the numerically calculated values of
the width to the parameter-free theoretical prediction (solid
line) based on RMT\cite{longpaper}. On average, the RMT based
prediction seems consistent with the numerics, though there is a
lot of variation in the widths driven by large variations in the
matrix elements coupling the quasiparticle levels to the
collective mode.

\section{Conclusions}
\medskip

In our earlier work, we used a global RG assumption to reduce the
problem on the scale of the Thouless energy to that of a disordered
noninteracting problem with Fermi-liquid interactions. This is quite
plausible for ballistic dots on very general grounds. To proceed
further we had to make two further assumptions; (i) That the $g$
approximate momentum states at the Fermi energy were a good basis in
which to expand the exact disorder eigenstates, and (ii) That the wave
functions of the exact eigenstates in the momentum basis obeyed all
the statistical properties of RMT. Based on these two assumptions we
were able to construct a solution to the problem which was
asymptotically exact in the limit $g\to\infty$. This solution led to
specific predictions for various physical quantities, including the
size of the order parameter, the reduction in energy due to
interactions, the shape of the energy landscape, and the size of the
quasiparticle decay widths.

In this paper we have carried out a calculation which is still
predicated on the validity of the Fermi-liquid form of the
interactions on a scale $E_L$ much larger than the Thouless energy
$E_T$. In retaining this assumption we are on firm ground, since after
all, the Thouless energy can be made as small as one wishes merely by
increasing the size of the system. We also assumed that the mean-field
description of the Landau Fermi-liquid interactions is valid, which is
justified by the fact that the minima in the effective potential
landscape are indeed of order $g^2\d$. However, we explicitly eschewed
the other two assumptions that we made in previous work, with a view
to independently testing their validity. We found that our first
assumption, that the approximate momentum states were a good basis in
which to expand the exact disorder eigenstates, was extremely good
near the Fermi energy, but became increasingly inaccurate as one went
to the edge of the Thouless shell. We did not test the second
assumption about wavefunction correlations explicitly, but indirectly
through its effects on the predictions of our earlier work. We found
that most of the predictions held up, with the exception of the shape
of the effective potential landscape in the case of symmetry-breaking
in the $m=2$ channel. Even here, the minimum is shaped more like a
crescent, indicating the possible emergence of the Mexican Hat
structure at larger values of $g$ (we went to the largest value of $g$
that we could given that we kept only 585 states and had to keep at
least half the states empty). We traced the discrepancy back to the
anomalously broad distribution (compared to estimates based on a
complex gaussian distribution) of the matrix elements
$M^x_{\a\b}$. However, we were unable to pin down a physical reason
for this broad distribution for the $m=2$ case.

In conclusion, much of the physics we uncovered using our RMT
assumptions seems to be valid in the Robnik-Berry billiard. The
second-order transition that we uncovered in the $g\to\infty$ limit
seems to indeed be broadened into a smooth croossover as expected,
belying fears that it may be overtaken by a first-order bulk
transition. The question of how large $g$ has to be before RMT becomes
fully applicable remains open; another way to phrase the question is
to ask what the nonuniversal corrections to RMT are in ballistic
systems. Finally, an important open question is whether the broad
distributions of matrix elements of interaction operators is a generic
feature of ballistic systems, rather than being a special feature of
the Robnik-Berry billiard, and if so, what physics determines the
width of those distributions. However, our results here give us
encouragement that the RMT assumptions can indeed be used with
confidence in making predictions in ballistic systems, at a
qualitative and semi-quantitative level.
\section{Acknowledgements}
We would like to thank Yoram Alhassid, Alex Barnett, Piet Brouwer,
and Doug Stone for illuminating conversations, and the Aspen
Center for Physics where part of this work was carried out. We are
also grateful to the NSF for partial support under grants DMR
0311761 (GM), DMR 0354517(RS).


\begin{thebibliography}{99}
\bibitem{recent-expts} U. Sivan {\it et al}, \prl\  {\bf 77}, 1123 (1996);
S. R. Patel {\it et al}, \prl\ {\bf 80}, 4522 (1998); F. Simmel {\it
et al}, \prb\ {\bf 59}, 10441 (1999); D. Abusch-Magder {\it et al},
Physica E\ {\bf 6}, 382 (2000); F. Simmel, T. Heinzel, and
D. A. Wharam, Eur. Lett. \ {\bf 38}, 123 (1997); J. A. Folk {\it et
al},
\prl\ {\bf 86}, 2102 (2001); S. L\"uscher {\it et al}, \prl\ {\bf 86}, 2118 (2001).
\bibitem{reviews} For recent reviews, see, T. Guhr, A. M\"uller-Groeling,
and H. A. Weidenm\"uller, Phys. Rep. {\bf 299}, 189 (1998);
Y. Alhassid, \rmp\ {\bf 72}, 895 (2000); A. D. Mirlin, Phys. Rep. {\bf
326}, 259 (2000).
\bibitem{H_U} A.  V.  Andreev and A.
Kamenev, \prl {\bf 81}, 3199 (1998); P.  W.  Brouwer, Y.  Oreg, and B.
I.  Halperin, \prb {\bf 60}, R13977 (1999); H.  U.  Baranger, D.
Ullmo, and L.  I.  Glazman, \prb {\bf 61}, R2425 (2000); I.  L.
Kurland, I.  L.  Aleiner, and B.  L.  Al'tshuler, \prb\ {\bf 62},
14886 (2000).
\bibitem{univ-ham}I. L. Aleiner, P. W. Brouwer, and L. I. Glazman,
Phys. Rep. {\bf 358}, 309 (2002), and references therein; Y. Oreg, P. W. Brouwer,
X. Waintal, and B. I. Halperin, cond-mat/0109541, and references
therein.
\bibitem{mm} G. Murthy and H. Mathur, \prl\ {\bf 89}, 126804 (2002).
\bibitem{qd-ms1} G. Murthy and R. Shankar, \prl {\bf 90}, 066801 (2003).
\bibitem{longpaper} G. Murthy, R. Shankar, D. Herman, and H. Mathur,
\prb {\bf 69}, 075321 (2004).
\bibitem{rg-shankar} R. Shankar, {\it Physica}\ {\bf A177},
530 (1991); R.Shankar, { Rev. Mod. Phys.} {\bf 66}, 129 (1994).
\bibitem{alt1} K. B. Efetov, Adv. Phys. {\bf 32}, 53 (1983);
B. L. Al'tshuler ad B. I. Shklovskii, { Sov. Phys. JETP}\
{\bf 64}, 127 (1986).
\bibitem{RMT} M. L. Mehta, {\it Random Matrices}, Academic Press, San
Diego, 1991.
\bibitem{critical-fan} S. Chakravarty, B. I. Halperin, and
D. R. Nelson, \prl\ {\bf 60}, 1057 (1988); \prb\ {\bf 39}, 2344
(1989); For a detailed treatment of the generality of the phenomenon,
see, S. Sachdev, {\it Quantum Phase Transitions}, Cambridge University
Press, Cambridge 1999.
\bibitem{robnik-berry} M. Robnik, J. Phys. A {\bf 17}, 1049 (1984);
M. V. Berry and M. Robnik, J. Phys. A {\bf 19}, 649 (1986).
\bibitem{agd} A. A.  Abrikosov, L. P. Gorkov, and I. E. Dzyaloshinski,
{\it Methods of Quantum Field Theory in Statistical Physics}, Dover
Publications, New York, 1963.
\bibitem{berry-ansatz} M. V. Berry, J. Phys. A {\bf 10}, 2083 (1977).
\bibitem{stone-bruus} A. D. Stone and H. Bruus, Physica B {\bf 189}, 43 (1993);
Surface Science {\bf 305}, 490 (1994).
\bibitem{alhassid-lewenkopf} Y. Alhassid and C. H. Lewenkopf, \prb {\bf 55}, 7749 (1997).
\bibitem{supersymm} K. B. Efetov and V. R. Kogan, \prb\ {\bf 67}, 245312 (2003), 
and references therein. 
\bibitem{nowindow} S. Adam, P. W. Brouwer, and P. Sharma,\prb {\bf 68}, 241311 (2003)
\bibitem{barnett-pvt} A. H. Barnett, private communication (2004).
\bibitem{barnett} A. H. Barnett, ``Asymptotic rate of quantum ergodicity in
chaotic Euclidean billiards'', Courant Institute preprint, 2004.
\end{thebibliography}
\end{document}